\documentclass[aps,prb,showpacs,reprint,a4paper]{revtex4-1}
\usepackage{graphicx}
\usepackage{bm}

\begin{document}

\title{Analytic evaluation of the electronic self-energy in the $\bm{GW}$ approximation\\ for two electrons on a sphere}
\author{Arno Schindlmayr}
\email{Arno.Schindlmayr@uni-paderborn.de}
\affiliation{Department Physik, Universit\"at Paderborn, 33095 Paderborn, Germany}
\affiliation{Institute for Solid State Physics, University of Tokyo, Kashiwa, Chiba 277-8581, Japan}

\date{5 February 2013}

\begin{abstract}
The $GW$ approximation for the electronic self-energy is an important tool for the quantitative prediction of excited states in solids, but its mathematical exploration is hampered by the fact that it must, in general, be evaluated numerically even for very simple systems. In this paper I describe a nontrivial model consisting of two electrons on the surface of a sphere, interacting with the normal long-range Coulomb potential, and show that the $GW$ self-energy, in the absence of self-consistency, can in fact be derived completely analytically in this case. The resulting expression is subsequently used to analyze the convergence of the energy gap between the highest occupied and the lowest unoccupied quasiparticle orbital with respect to the total number of states included in the spectral summations. The asymptotic formula for the truncation error obtained in this way, whose dominant contribution is proportional to the cutoff energy to the power $-3/2$, may be adapted to extrapolate energy gaps in other systems.
\end{abstract}


\maketitle

\section{Introduction}

Accurate first-principles calculations of electronic excitations in solids are notoriously challenging, because the variational principle, which underlies ground-state schemes like Kohn-Sham density-functional theory\cite{Hohenberg1964,Kohn1965} or quantum Monte Carlo methods,\cite{Foulkes2001} cannot be exploited in this case. Therefore, many-body perturbation theory,\cite{Mahan1990} which is based on Green functions and allows an explicit incorporation of relevant Coulomb-correlation effects through a summation of the corresponding Feynman diagrams, is often the method of choice for quantitative \textit{ab initio} descriptions of experimental spectroscopies. An especially fruitful realization of this framework for actual electronic-structure calculations is the so-called $GW$ approximation\cite{Hedin1965} that yields quasiparticle band structures in much better agreement with photoemission data than density-functional theory with standard local or semilocal exchange-correlation functionals.\cite{Aulbur2000} From a mathematical point of view, it constitutes an expansion of the exact nonlocal and frequency-dependent self-energy to first order in the dynamically screened Coulomb potential $W$, which describes the interaction between two quasiparticles formed by an electron or a hole together with its surrounding polarization cloud. The $GW$ approximation is hence particularly suited for materials with weak to medium correlation strength, such as semiconductors or simple metals, but not for strongly correlated systems, where the quasiparticle picture breaks down.

In spite of the undisputed success of the $GW$ approximation for the prediction of material properties and the interpretation of photoemission measurements, the debate over its best practical implementation shows no tendency of abating. One long-standing controversy centers on the question whether the Green function used to construct the self-energy should be evaluated self-consistently or not.\cite{Shirley1996,Holm1998,Schone1998,Schindlmayr1998b,Takada2001,Ku2002,Faleev2004,Bruneval2006,Shishkin2007a,Shishkin2007b,Stan2009} Both approaches can be justified: The Green function obtained with full self-consistency is the variational solution that makes the Luttinger-Ward functional\cite{Luttinger1960} for the total energy within the random-phase approximation stationary,\cite{Baym1962} while the more commonly applied non-self-consistent version follows naturally from the iterative solution of Hedin's coupled integral equations for the self-energy with a mean-field treatment as the starting point.\cite{Hedin1965}

Assessments of different variants of the $GW$ approximation are typically based on a comparison of the numerical results for selected test systems. However, the adopted numerical procedures and limited computational resources inevitably necessitate additional simplifications whose impact on the final results is not always clear. For instance, the case studies investigating the effects of self-consistency for real solids reported so far have, in general, only used a restricted form of self-consistency that was limited to the quasiparticle energies in the denominator of the Green function,\cite{Shishkin2007a} the quasiparticle orbitals,\cite{Faleev2004} or the diagonal part of the Green function.\cite{Ku2002} Even at the level of the standard non-self-consistent $GW$ approximation, far-reaching additional simplifications are widely employed. Among these are the pseudopotential approximation, which leads to small but systematic deviations from all-electron results due to the inexact core-valence partitioning and the use of pseudo wave functions,\cite{Kotani2002,Shishkin2006,Friedrich2006,Gomez-Abal2008,Friedrich2010,Li2012} as well as plasmon-pole models or other simplified screening functions.\cite{Aulbur2000} A range of less obvious but equally important factors like the proper treatment of anisotropic screening in noncubic systems\cite{Kotani2002,Freysoldt2007,Friedrich2009} or problems resulting from an incomplete basis set in all-electron calculations\cite{Friedrich2006,vanSchilfgaarde2006} have also been emphasized.

One point that has increasingly come into focus in this context is the convergence behavior with respect to the number of unoccupied states included in the construction of the Green function $G$ and the screened Coulomb interaction $W$. Much attention was raised after an early all-electron calculation,\cite{Ku2002} which claimed much smaller semiconductor band gaps than established pseudopotential results, was put into question due to alleged incomplete convergence with respect to this parameter,\cite{Tiago2004,Delaney2004} leading to further detailed studies\cite{Friedrich2006,vanSchilfgaarde2006,Bruneval2008} that illustrated the slow convergence rate of the quasiparticle energies for a wide variety of materials. As an extreme example, Shih \textit{et al.}\cite{Shih2010} reported that bulk zinc oxide required thousands of unoccupied bands to achieve satisfactory accuracy in a calculation based on pseudopotentials and a plane-wave basis set. It was subsequently argued that this peculiar behavior was caused by the particular choice of plasmon-pole model used in Ref.\ \onlinecite{Shih2010} and that far fewer bands are in fact required if the full frequency-dependent screening function is properly constructed within the random-phase approximation.\cite{Stankovski2011} However, a similarly slow convergence was again observed in an all-electron calculation for zinc oxide that not only avoided plasmon-pole models but also the additional pseudopotential approximation.\cite{Friedrich2011} 
Parallel to these developments, different approaches were proposed to circumvent or at least alleviate the convergence problem. These include the replacement of all high-lying empty states by plane waves,\cite{Steinbeck2000} the extrapolar method,\cite{Bruneval2008} in which merely a small number of unoccupied states are treated explicitly and a common energy denominator is assigned to the remainder, so that the closure relation can be applied, the Lanczos-chain algorithm,\cite{Umari2010} as well as the effective-energy technique\cite{Berger2010} and methods based on the self-consistent Sternheimer equation,\cite{Giustino2010} which are formally exact despite only involving occupied states. While these novel schemes undoubtedly hold great potential, practical applications are not yet widespread, in part because not all popular computer codes support them at present. As a consequence, the majority of $GW$ calculations still rely on traditional procedures and suffer from the problem of slow convergence.

All of the above issues are relevant on the energy scale of several tenths of an electron volt that matters for the comparison between different implementations of the $GW$ approximation and with experiments, but their control is difficult in practice due to complex interdependencies. Therefore, model systems that permit numerically exact or, ideally, analytic solutions play an important role for developing and testing approximation schemes within many-body perturbation theory, but even the homogeneous electron gas, a frequently employed model in solid-state physics, can only be treated numerically in the $GW$ approximation. Furthermore, with no experimental measurements or independent theoretical benchmark results, even the basic question whether the true occupied band width in the range of metallic densities is smaller than that of free electrons, as predicted by the standard non-self-consistent $GW$ approximation,\cite{Hedin1965,Mahan1989} or larger, as obtained when full self-consistency is included,\cite{Shirley1996,Holm1998} is not yet finally settled. Calculations that go beyond the $GW$ approximation and attempt to incorporate the combined effects of self-consistency and vertex corrections remain inconclusive, because the results depend on the choice of vertex function and details of the implementation.\cite{Shirley1996,Takada2001,Holm1997} In this situation, more tractable few-electron systems are of considerable interest.

The first nontrivial system for which the self-energy at the $GW$ level can be derived analytically was a Hubbard model with four sites in a tetrahedral arrangement and two electrons,\cite{Schindlmayr1997} originally used as a counterexample to demonstrate the violation of particle-number conservation in the non-self-consistent $GW$ approximation before a more general investigation of this problem based on symmetry arguments.\cite{Schindlmayr2001} The analytic solvability was important in this case, because it proved unequivocally that the quantitative deviation was genuine and not due to numerical inaccuracies. A related but even simpler two-site model with a pair of electrons can be treated analytically in the same way.\cite{Romaniello2009} Lattice models with a wider range of parameters, for which the $GW$ self-energy is accurately obtainable by numerical means, were also employed in several studies.\cite{Schindlmayr1998b,Verdozzi1995,Pollehn1998,Sun2004,Kaasbjerg2010} The properties of Hubbard models deviate in many respects from those of real materials, however, and conclusions from such comparisons cannot always be directly transferred to the \textit{ab initio} realm.\cite{Pollehn1998} Most importantly, the local on-site interaction differs significantly from the actual Coulomb potential and leads to a dominance of short-range correlation effects. These are not well described by the $GW$ approximation, which mainly accounts for the long-range screening of charge carriers. Furthermore, the restricted Hilbert space does not allow us to address problems like the convergence behavior with respect to the number of empty states.\cite{Shih2010} Peculiar symmetries, such as that between occupied and unoccupied states in the two-site model at half filling,\cite{Romaniello2009} which are not obeyed by real solids, may also have an influence on the results. For completeness, it should be mentioned that the polaron model of individual electrons coupled to an external boson field can also be treated analytically,\cite{Hedin1999} but its usefulness as a test system for the $GW$ approximation is even more limited, as there is no explicit renormalizable electron-electron interaction.

For future methodological investigations I here propose a better suited continuum system consisting of two electrons confined to the surface of a sphere, and I show that the self-energy within the standard non-self-consistent $GW$ approximation can be derived entirely analytically. In contrast to the previously considered Hubbard models with the same property,\cite{Schindlmayr1997,Romaniello2009} the electrons interact with the normal long-range Coulomb potential, and there is an infinite Hilbert space of single-particle wave functions whose eigenvalues are not bounded from above. Therefore, the performance of particular approximation schemes should be more indicative of applications to real systems. The model considered here can be regarded as a two-dimensional homogeneous electron gas in a closed curved space, whose density depends on the radius of the sphere. As in the three-dimensional electron gas, the correlation is weak at high densities (small radius) and becomes strong at low densities (large radius),\cite{Loos2009a} so that different regimes can be explored within the same framework. The system is also quasi-exactly solvable, which means that some exact eigenvalues, although not the complete spectrum, are known analytically.\cite{Loos2009b}

This paper is organized as follows: In Sec.\ \ref{Sec:model} the system is mathematically defined and discussed in more detail, before the analytic expression for the self-energy in the non-self-consistent $GW$ approximation is derived in Sec.\ \ref{Sec:self-energy}. Then in Sec.\ \ref{Sec:convergence} the convergence behavior with respect to the number of empty states is analyzed, leading to an analytic formula describing the asymptotic dependence on the cutoff energy. The conclusions are summarized in Sec.\ \ref{Sec:conclusions}. Unless otherwise noted, Hartree atomic units are used throughout.

\section{Model description\label{Sec:model}}

The system considered here consists of two electrons on the two-dimensional surface of a sphere with radius $R$. Their positions are expressed in spherical coordinates
\begin{equation}
\mathbf{r}(\theta,\phi) = R \left(
\begin{array}{c}
\sin \theta \cos \phi \\
\sin \theta \sin \phi \\
\cos \theta
\end{array}
\right)
\end{equation}
in terms of the polar angle $\theta$ and the azimuthal angle $\phi$. A homogeneous positive surface charge density $2/(4 \pi R^2)$, which gives rise to the attractive electrostatic potential $-2 / R$, is included to ensure overall charge neutrality. If the self-interaction of the positive charge background is also taken into account, then the Hamiltonian becomes
\begin{equation}\label{Eq:Hamiltonian-2}
H = -\frac{\Delta_{S^2}}{2 R^2} - \frac{\Delta'_{S^2}}{2 R^2} + \frac{1}{|\mathbf{r}(\theta,\phi) - \mathbf{r}(\theta',\phi')|} - \frac{2}{R}
\end{equation}
with the Laplace-Beltrami operator on the 2-sphere
\begin{equation}
\Delta_{S^2} = \frac{1}{\sin \theta} \frac{\partial}{\partial \theta} \sin \theta \frac{\partial}{\partial \theta} + \frac{1}{\sin^2 \theta} \frac{\partial^2}{\partial \phi^2} \;,
\end{equation}
which is related to the angular-momentum operator $\mathbf{L}$. Its eigenfunctions are the spherical harmonics
\begin{equation}\label{Eq:angular-momentum}
-\Delta_{S^2} Y_{\ell m}(\theta,\phi) = \mathbf{L}^2 Y_{\ell m}(\theta,\phi) = \ell (\ell + 1) Y_{\ell m}(\theta,\phi) \;.
\end{equation}
The eigenstates of the Hamiltonian (\ref{Eq:Hamiltonian-2}) must in general be determined numerically, but for certain discrete radii individual analytic solutions are known.\cite{Loos2009b} For example, for the particular value $R = \sqrt{3} / 2$, which corresponds to intermediate correlation strength, the exact ground-state energy for two electrons is $E_0(2) = 1 - 4 / \sqrt{3}$.

Due to symmetry requirements, the ground-state electron density is evenly distributed on the spherical surface. In density-functional theory the one-particle Hamiltonian of the auxiliary Kohn-Sham system thus takes the form
\begin{equation}
h = -\frac{\Delta_{S^2}}{2 R^2} + V_\mathrm{xc}
\end{equation}
with a constant exchange-correlation potential $V_\mathrm{xc}$. The Hartree potential is exactly canceled by the electrostatic potential of the positive charge background. From Eq.\ (\ref{Eq:angular-momentum}) the Kohn-Sham orbitals are
\begin{equation}\label{Eq:KS-orbitals}
y_{\ell m}(\theta,\phi) = \frac{Y_{\ell m}(\theta,\phi)}{R} \;,
\end{equation}
where the normalization is chosen with respect to the two-dimensional integral over the spherical surface
\begin{equation}
\int_0^{2 \pi} \int_0^\pi y_{\ell m}^*(\theta,\phi) y_{\ell' m'}(\theta,\phi) R^2 \sin \theta \,d\theta \,d\phi = \delta_{\ell \ell'} \delta_{m m'} \;,
\end{equation}
and the corresponding eigenvalues
\begin{equation}\label{Eq:KS-eigenvalues}
\epsilon_\ell = \frac{\ell (\ell + 1)}{2 R^2} + V_\mathrm{xc}
\end{equation}
are independent of the magnetic quantum number $m$. In the ground state the lowest spin-degenerate Kohn-Sham orbital is doubly occupied, whereas all others are unoccupied. As the eigenvalue of the highest occupied orbital also equals the negative of the ionization potential in exact density-functional theory,\cite{Levy1984} the exchange-correlation potential can be determined from the difference
\begin{equation}
V_\mathrm{xc} = \epsilon_0 = E_0(2) - E_0(1)
\end{equation}
between the ground-state total energy $E_0(2)$ of the true interacting two-electron system and the energy $E_0(1) = 0$ of the corresponding ionized one-electron system, if the former is known. Thus for $R = \sqrt{3}/2$ the exact exchange-correlation potential is $V_\mathrm{xc} = 1 - 4 / \sqrt{3}$. Alternatively, for free electrons $V_\mathrm{xc}$ is set to zero. Analogous to the three-dimensional homogeneous electron gas, these two choices differ only by a trivial energy shift.

\section{Derivation of the self-energy\label{Sec:self-energy}}

The first ingredient required for the construction of the $GW$ self-energy is the Kohn-Sham Green function. In the following, all functions will be projected onto the orbitals (\ref{Eq:KS-orbitals}), which form a complete set. The Green function is diagonal in this basis, and the diagonal matrix elements
\begin{equation}
G_\ell(\omega) = \frac{\delta_{\ell 0}}{\omega - \epsilon_0 - i \eta} + \frac{1 - \delta_{\ell 0}}{\omega - \epsilon_\ell + i \eta}
\end{equation}
are furthermore independent of $m$. The symbol $\eta$ always denotes a positive infinitesimal. Since the wave function $y_{00} = (4 \pi R^2)^{-1/2}$ of the only occupied Kohn-Sham state is constant, the polarization function becomes
\begin{eqnarray}
P_\ell(\omega) &=& -2 | y_{00} |^2 \frac{i}{2 \pi} \int_{-\infty}^\infty \left[ G_\ell(\omega+\omega') G_0(\omega') \right. \nonumber\\
&&\left. \mbox{} + G_0(\omega+\omega') G_\ell(\omega') \right] \,d\omega' \\
&=& \frac{1 - \delta_{\ell 0}}{2 \pi R^2} \left( \frac{1}{\omega - \omega_\ell + i \eta} - \frac{1}{\omega + \omega_\ell - i \eta} \right) \nonumber
\end{eqnarray}
with the definition $\omega_\ell = \epsilon_\ell - \epsilon_0$ and a factor 2 for the spin summation. The representation of the Coulomb potential
\begin{equation}
v_\ell = \frac{4 \pi}{2 \ell + 1} R
\end{equation}
follows from the addition theorem for the spherical harmonics. It is immediately clear from this formula that the interaction strength grows with $R$ while the level spacing between the eigenvalues (\ref{Eq:KS-eigenvalues}) is simultaneously reduced, so that the system becomes more and more strongly correlated with increasing sphere radius. As the polarization function and the Coulomb potential are both diagonal in the chosen basis, the matrix elements of the dynamically screened interaction
\begin{equation}
W_\ell(\omega) = v_\ell + W^\mathrm{c}_\ell(\omega)
\end{equation}
with the correlation part
\begin{eqnarray}
W^\mathrm{c}_\ell(\omega) &=& v_\ell \frac{P_\ell(\omega)}{1 - v_\ell P_\ell(\omega)} v_\ell \\
&=& \frac{(1 - \delta_{\ell 0}) 8 \pi \omega_\ell}{(2 \ell + 1)^2 z_\ell} \left( \frac{1}{\omega - z_\ell + i \eta} - \frac{1}{\omega + z_\ell - i \eta} \right) \nonumber
\end{eqnarray}
can be calculated by means of a simple scalar renormalization. The poles are located at
\begin{equation}
z_\ell = \sqrt{ \omega_\ell^2 + \frac{4 \omega_\ell}{(2 \ell + 1) R} } \;.
\end{equation}
The exchange part of the self-energy, which is frequency independent and equals the nonlocal exchange potential in Hartree-Fock theory, is obtained as
\begin{equation}
\Sigma^\mathrm{x}_\ell = -| y_{00} |^2 v_\ell = -\frac{1}{(2 \ell + 1) R} \;,
\end{equation}
while the correlation part of the self-energy is given by a convolution of the Green function and the screened interaction. The matrix elements are given by
\begin{eqnarray}
\Sigma^\mathrm{c}_{\ell m, \ell' m'}(\omega) &=& \sum_{\ell_1 = 0}^\infty \sum_{m_1 = -\ell_1}^{\ell_1} \sum_{\ell_2 = 0}^\infty \sum_{m_2 = -\ell_2}^{\ell_2} \frac{1}{R^2} \\
&& \mbox{} \times  \langle \ell_1 m_1, \ell_2 m_2; \ell m \rangle \langle \ell_1 m_1, \ell_2 m_2; \ell' m' \rangle^* \nonumber\\
&& \mbox{} \times  \frac{i}{2 \pi} \int_{-\infty}^\infty G_{\ell_1}(\omega - \omega') W^\mathrm{c}_{\ell_2}(\omega') \,d\omega' \nonumber
\end{eqnarray}
with the Gaunt coefficients\cite{Gaunt1929}
\begin{eqnarray}
\langle \ell_1 m_1, \ell_2 m_2; \ell m \rangle &=& \int_0^{2 \pi} \int_0^\pi Y_{\ell_1 m_1}(\theta,\phi) Y_{\ell_2 m_2}(\theta,\phi) \nonumber\\
&&\mbox{} \times Y_{\ell m}^*(\theta,\phi) \sin \theta \,d\theta \,d\phi \;.
\end{eqnarray}
These overlap integrals are in fact real valued and zero unless $m_1 + m_2 = m$. They are written conveniently in terms of the Wigner 3-$j$ symbols\cite{Messiah1962} as
\begin{eqnarray}
\langle \ell_1 m_1, \ell_2 m_2; \ell m \rangle &=& (-1)^m \sqrt{ \frac{(2 \ell_1 + 1) (2 \ell_2 + 1) (2 \ell + 1)}{4 \pi} } \nonumber \\
&& \mbox{} \times \left(
\begin{array}{ccc}
\ell_1 & \ell_2 & \ell \\
0 & 0 & 0
\end{array}
\right) \left(
\begin{array}{ccc}
\ell_1 & \ell_2 & \ell \\
m_1 & m_2 & -m
\end{array}
\right) \;. \nonumber \\
\end{eqnarray}
It then follows from the orthogonality relation
\begin{equation}
\sum_{m_1 = -\ell_1}^{\ell_1} \sum_{m_2 = -\ell_2}^{\ell_2} \left(
\begin{array}{ccc}
\ell_1 & \ell_2 & \ell \\
m_1 & m_2 & m
\end{array}
\right) \left(
\begin{array}{ccc}
\ell_1 & \ell_2 & \ell' \\
m_1 & m_2 & m'
\end{array}
\right)
= \frac{\delta_{\ell \ell'} \delta_{m m'}}{2 \ell + 1}
\end{equation}
that the correlation part of the self-energy, like all other quantities considered in this section, is diagonal in the Kohn-Sham basis and that the diagonal elements
\begin{eqnarray}
\Sigma^\mathrm{c}_\ell(\omega) &=& \sum_{\ell_1 = 0}^\infty \sum_{\ell_2 = 0}^\infty \frac{(2 \ell_1 + 1) (2 \ell_2 + 1)}{4 \pi R^2} \left(
\begin{array}{ccc}
\ell_1 & \ell_2 & \ell \\
0 & 0 & 0
\end{array}
\right)^2 \nonumber \\
&& \mbox{} \times \frac{i}{2 \pi} \int_{-\infty}^\infty G_{\ell_1}(\omega - \omega') W^\mathrm{c}_{\ell_2}(\omega') \,d\omega'
\end{eqnarray}
are again independent of $m$. Carrying out the remaining contour integration eventually yields
\begin{eqnarray}
\Sigma^\mathrm{c}_\ell(\omega) &=& \frac{(1 - \delta_{\ell 0}) 2 \omega_\ell}{(2 \ell + 1)^2 z_\ell R^2} \frac{1}{\omega - \epsilon_0 + z_\ell - i \eta} \nonumber \\
&& \mbox{} + \sum_{\ell_1 = 1}^\infty \sum_{\ell_2 = 1}^\infty \left(
\begin{array}{ccc}
\ell_1 & \ell_2 & \ell \\
0 & 0 & 0
\end{array}
\right)^2 \label{Eq:self-energy} \\
&& \mbox{} \times \frac{2 (2 \ell_1 + 1) \omega_{\ell_2}}{(2 \ell_2 + 1) z_{\ell_2} R^2} \frac{1}{\omega - \epsilon_{\ell_1} - z_{\ell_2} + i \eta} \;. \nonumber
\end{eqnarray}

\begin{figure}
\includegraphics[clip,width=\linewidth]{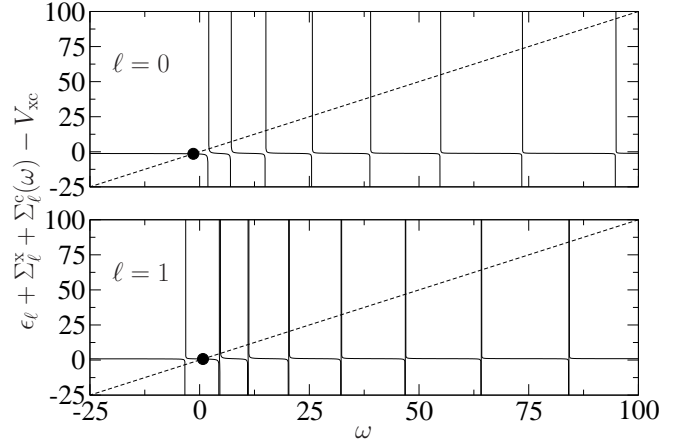}
\caption{Right-hand side $\epsilon_\ell + \Sigma^\mathrm{x}_\ell + \Sigma^\mathrm{c}_\ell(\omega) - V_\mathrm{xc}$ of the quasiparticle equation (\ref{Eq:quasiparticle}) for the occupied state ($\ell = 0$, above) and the first unoccupied state ($\ell = 1$, below) for $R = \sqrt{3}/2$. The solutions of the quasiparticle equation correspond to the intersections with the diagonal $\omega$ (dashed line). In addition to the actual quasiparticle, marked by a circle, there are an infinite number of satellite resonances for each value of $\ell$.\label{Fig:Sigma}}
\end{figure}

This general formula can now be exploited to derive the self-energy correction of individual orbitals. Of particular interest are the highest (and only) occupied state
\begin{equation}\label{Eq:Sigma-0}
\Sigma^\mathrm{c}_0(\omega) = \sum_{\ell = 1}^\infty \frac{2 \omega_\ell}{(2 \ell + 1) z_\ell R^2} \frac{1}{\omega - \epsilon_\ell - z_\ell + i \eta}
\end{equation}
as well as the lowest unoccupied state
\begin{eqnarray}
\Sigma^\mathrm{c}_1(\omega) &=& \frac{2 \omega_1}{9 z_1 R^2} \frac{1}{\omega - \epsilon_0 + z_1 - i \eta} \label{Eq:Sigma-1} \\
&& \mbox{} + \sum_{\ell = 2}^\infty \frac{2 \ell \omega_{\ell - 1}}{(2 \ell - 1)^2 z_{\ell - 1} R^2} \frac{1}{\omega - \epsilon_\ell - z_{\ell - 1} + i \eta} \nonumber\\
&& \mbox{} + \sum_{\ell = 2}^\infty \frac{2 \ell \omega_\ell}{(2 \ell + 1)^2 z_\ell R^2} \frac{1}{\omega - \epsilon_{\ell - 1} - z_\ell + i \eta} \;. \nonumber
\end{eqnarray}
Finally, the energy spectrum of electronic excitations is given by the solutions of the quasiparticle equation
\begin{equation}\label{Eq:quasiparticle}
\omega = \epsilon_\ell + \Sigma^\mathrm{x}_\ell + \Sigma^\mathrm{c}_\ell(\omega) - V_\mathrm{xc} \;.
\end{equation}
In Fig.\ \ref{Fig:Sigma} both sides of this nonlinear equation are shown for $\ell = 0$ (above) and $\ell = 1$ (below) for a sphere radius of $R = \sqrt{3}/2$. The possible excitations correspond to the intersections of the two curves. Evidently, there are an infinite number of satellite resonances for each quantum number $\ell$, stemming from the poles of the self-energy, in addition to the principal quasiparticle state, which is marked by a circle. It should be noted that the poles are well separated at the positions $\epsilon_\ell + z_\ell$ with integer $\ell$ for the occupied state, while the self-energy of the unoccupied state features pairs of poles located very close to each other at $\epsilon_\ell + z_{\ell - 1}$ and $\epsilon_{\ell - 1} + z_\ell$. These stem from the second and third term on the right-hand side of Eq.\ (\ref{Eq:Sigma-1}), but the splitting cannot be properly resolved on the scale of the figure. As a rule, the multiplicity of the satellites splitting increases with the quantum number of the quasiparticle state. If the self-energy is linearized around the Kohn-Sham eigenvalues $\epsilon_\ell$, as is common in \textit{ab initio} calculations, then the quasiparticle energies are given by
\begin{equation}\label{Eq:linearization}
\epsilon^\mathrm{qp}_\ell = \epsilon_\ell + Z_\ell \left[ \Sigma^\mathrm{x}_\ell + \Sigma^\mathrm{c}_\ell(\epsilon_\ell) - V_\mathrm{xc} \right] \;,
\end{equation}
where the renormalization factors
\begin{equation}\label{Eq:renormalization}
Z_\ell = \left. \frac{1}{1 - \frac{\partial}{\partial \omega} \Sigma^\mathrm{c}_\ell(\omega)} \right|_{\omega = \epsilon_\ell}
\end{equation}
specify the weight of the quasiparticle resonance in the spectral function. For $R = \sqrt{3}/2$ the values are $Z_0 \approx 0.94$ and $Z_1 \approx 0.95$.

\begin{figure}
\includegraphics[clip,width=\linewidth]{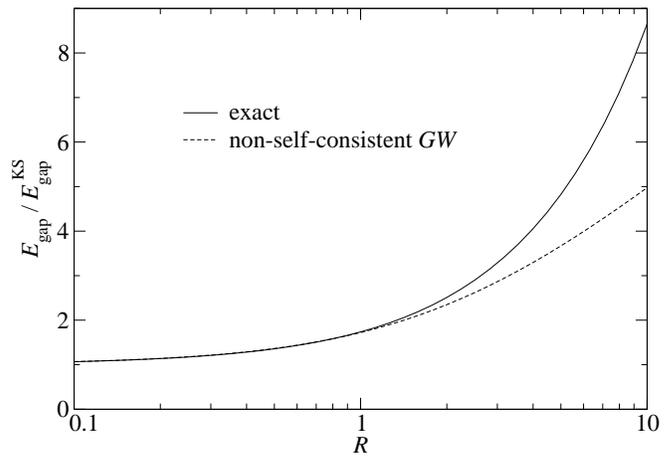}
\caption{Energy gap $E_\mathrm{gap} = \epsilon^\mathrm{qp}_1 - \epsilon^\mathrm{qp}_0$ between the occupied and the lowest unoccupied quasiparticle state as a function of the sphere radius $R$ relative to the Kohn-Sham eigenvalue gap $E_\mathrm{gap}^\mathrm{KS} = \epsilon_1 - \epsilon_0$ (dashed line). The exact numerical value, obtained from the difference between the ionization potential and the electron affinity, is also shown (solid line). Small $R$ correspond to weak and large $R$ to strong correlation.\label{Fig:energygap}}
\end{figure}

The energy gap $E_\mathrm{gap} = \epsilon^\mathrm{qp}_1 - \epsilon^\mathrm{qp}_0$ thus obtained is shown in Fig.\ \ref{Fig:energygap} as a function of the sphere radius $R$ relative to the Kohn-Sham eigenvalue gap $E_\mathrm{gap}^\mathrm{KS} = \epsilon_1 - \epsilon_0$. In addition, the exact value, which is defined as the difference $I - A$ between the ionization potential $I = E_0(1) - E_0(2)$ and the electron affinity $A = E_0(2) - E_0(3)$, is displayed for comparison. The ground-state energies $E_0(2)$ for two electrons and $E_0(3)$ for three electrons are obtained from a numerical diagonalization with a full set of Slater determinants constructed from the orbitals (\ref{Eq:KS-orbitals}), converged at the scale of the figure. For small radii, i.e., high electron densities, the dynamics of the system are dominated by the kinetic energy, and the quasiparticle energy gap approaches that of noninteracting electrons. With increasing sphere radius, correlation effects become stronger, and the quasiparticle energy gap widens relative to the eigenvalue gap of the noninteracting Kohn-Sham system. Up to an intermediate correlation strength of $R \approx 1$, the $GW$ approximation is in excellent quantitative agreement with the exact numerical value, but it underestimates the further rapid increase of the gap in the strong-correlation regime at larger radii. Altogether, the behavior of the $GW$ approximation for this model system hence accords completely with its performance for real solids.

\section{Asymptotic convergence\label{Sec:convergence}}

The analytic expression (\ref{Eq:self-energy}) for the self-energy derived above includes a double infinite summation over the angular quantum numbers $\ell_1$ and $\ell_2$, reflecting the spectral sums over unoccupied eigenstates in the Green function and the screened interaction, respectively. In practice, such sums must be truncated at a certain cutoff energy, so that quantitative deviations from the true results are incurred as a consequence. In the following I analyze the asymptotic convergence with respect to the number of unoccupied states for this system.

Owing to the truncation, the exact matrix element of the correlation part of the self-energy in Eq.\ (\ref{Eq:linearization}) is replaced by $\tilde{\Sigma}^\mathrm{c}_\ell(\epsilon_\ell)$, which includes only those terms of  Eq.\ (\ref{Eq:self-energy}) where both $\ell_1$ and $\ell_2$ are smaller than a particular finite quantum number $L_\mathrm{cut}$. If the difference is denoted by $\Delta_\ell = \tilde{\Sigma}^\mathrm{c}_\ell(\epsilon_\ell) - \Sigma^\mathrm{c}_\ell(\epsilon_\ell)$, then the associated error in the quasiparticle energies due to the truncation equals $Z_\ell \Delta_\ell$. In principle, there is also another distinct error that stems from the approximate evaluation of the renormalization factors (\ref{Eq:renormalization}), which are in practice obtained from the truncated $\tilde{\Sigma}^\mathrm{c}_\ell(\omega)$ instead of the exact $\Sigma^\mathrm{c}_\ell(\omega)$, but it turns out that the resulting additional deviation is proportional to the cutoff energy to the power $-2$ and hence not relevant for the following analysis of the leading-order corrections.

In accordance with Eq.\ (\ref{Eq:Sigma-0}), the truncation changes the self-energy matrix element for the occupied state by
\begin{equation}
\Delta_0 = -\sum_{\ell = L_\mathrm{cut}}^\infty \frac{2 \omega_\ell}{(2 \ell + 1) z_\ell R^2} \frac{1}{\epsilon_0 - \epsilon_\ell - z_\ell} \;.
\end{equation}
If the addends on the right-hand side are expanded in inverse powers of $\ell$ according to
\begin{equation}
\Delta_0 = \sum_{\ell = L_\mathrm{cut}}^\infty \left( \frac{1}{\ell^3} - \frac{3}{2 \ell^4} + O(\ell^{-5}) \right) \;,
\end{equation}
then the entire expression can be rewritten as a sum of Hurwitz zeta functions
\begin{equation}
\zeta(n,L_\mathrm{cut}) = \sum_{\ell = L_\mathrm{cut}}^\infty \frac{1}{\ell^n} = \sum_{\ell = 0}^\infty \frac{1}{(L_\mathrm{cut} + \ell)^n}
\end{equation}
with positive integer exponents $n$. From the asymptotic behavior of the Hurwitz zeta function\cite{Erdelyi1953}
\begin{equation}
\zeta(n,L_\mathrm{cut}) = \frac{1}{(n - 1) L_\mathrm{cut}^{n-1}} + \frac{1}{2 L_\mathrm{cut}^n} + O(L_\mathrm{cut}^{-(n+1)})
\end{equation}
one thus obtains
\begin{equation}\label{Eq:Delta-0}
\Delta_0 = \frac{1}{2 L_\mathrm{cut}^2} + O(L_\mathrm{cut}^{-4}) = \frac{1}{4 R^2 E_\mathrm{cut}} + O(E_\mathrm{cut}^{-2}) \;.
\end{equation}
In the last step a cutoff energy
\begin{equation}\label{Eq:Ecut}
E_\mathrm{cut} = \frac{L_\mathrm{cut}^2}{2 R^2} \;,
\end{equation}
which is half-way between the energies of the states with $L_\mathrm{cut} - 1$ and $L_\mathrm{cut}$, was inserted. An analogous calculation for the first excited state based on Eq.\ (\ref{Eq:Sigma-1}) yields
\begin{equation}\label{Eq:Delta-1}
\Delta_1 = \frac{1}{4 R^2 E_\mathrm{cut}} + \frac{1}{4 \sqrt{2} R^3 E_\mathrm{cut}^{3/2}} + O(E_\mathrm{cut}^{-2}) \;. \nonumber
\end{equation}
As the terms proportional to $L_\mathrm{cut}^{-3}$ fail to fortuitously cancel in this case, the final expression retains a nonvanishing contribution with the cutoff energy to the power $-3/2$. The truncation error of the energy gap between the occupied and the lowest unoccupied quasiparticle state hence exhibits the asymptotic behavior
\begin{equation}\label{Eq:Delta-gap}
\Delta_\mathrm{gap} = Z_1 \Delta_1 - Z_0 \Delta_0 \sim \frac{Z_1 - Z_0}{4 R^2 E_\mathrm{cut}} + \frac{Z_1}{4 \sqrt{2} R^3 E_\mathrm{cut}^{3/2}} \;.
\end{equation}
The leading term is proportional to $E_\mathrm{cut}^{-1}$ but of small absolute magnitude, because the renormalization factors (\ref{Eq:renormalization}) in the vicinity of the fundamental gap show very little variation. In fact, if the quasiparticle energies are evaluated with $Z_\ell = 1$, as has been advocated by some authors,\cite{vanSchilfgaarde2006} then this term vanishes exactly. In practice, the convergence of the gap is hence dominated by the term proportional to $E_\mathrm{cut}^{-3/2}$. This remains true for other transitions; for example, between the occupied state and the second unoccupied state, whose truncation error
\begin{equation}
\Delta_2 = \frac{1}{4 R^2 E_\mathrm{cut}} + \frac{3}{8 \sqrt{2} R^3 E_\mathrm{cut}^{3/2}} + O(E_\mathrm{cut}^{-2})
\end{equation}
can be derived along the same lines. If the cutoff energy is not chosen as in Eq.\ (\ref{Eq:Ecut}) in the center but elsewhere in the interval between the states with quantum numbers $L_\mathrm{cut} - 1$ and $L_\mathrm{cut}$, then the formulas for all $\Delta_\ell$ are modified by an additional identical term proportional to $E_\mathrm{cut}^{-3/2}$.

\begin{figure}
\includegraphics[clip,width=\linewidth]{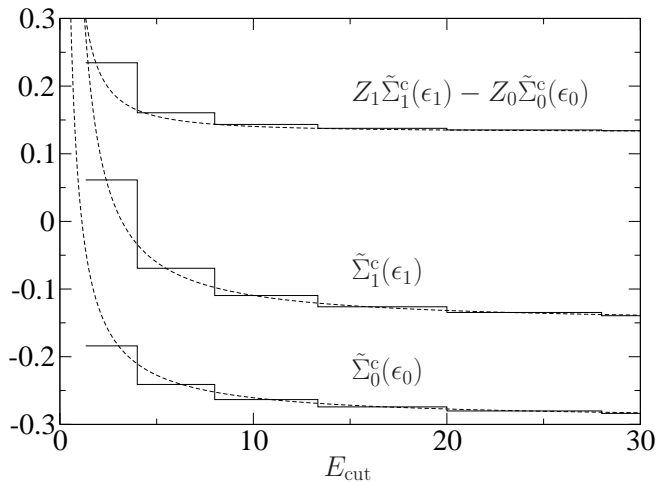}
\caption{Convergence of the matrix elements of the correlation part of the self-energy $\tilde{\Sigma}^\mathrm{c}_\ell(\epsilon_\ell)$ for the occupied state ($\ell = 0$) and the first unoccupied state ($\ell = 1$) as well as their contribution $Z_1 \tilde{\Sigma}^\mathrm{c}_1(\epsilon_1) - Z_0 \tilde{\Sigma}^\mathrm{c}_0(\epsilon_0)$ to the energy gap for $R = \sqrt{3}/2$. The dashed lines indicate the asymptotic behavior given by the analytic formulas (\ref{Eq:Delta-0}), (\ref{Eq:Delta-1}), and (\ref{Eq:Delta-gap}). In the latter case, only the term proportional to $E_\mathrm{cut}^{-3/2}$ is included.\label{Fig:convergence}}
\end{figure}

As a visualization, Fig.\ \ref{Fig:convergence} displays the matrix elements $\tilde{\Sigma}^\mathrm{c}_\ell(\epsilon_\ell)$ for the occupied state ($\ell = 0$) and for the first unoccupied state ($\ell = 1$) as a function of the cutoff energy for $R = \sqrt{3}/2$. The steplike variation reflects the discrete nature of the Kohn-Sham eigenvalue spectrum. The dashed lines indicate the asymptotic expansions $\Sigma^\mathrm{c}_\ell(\epsilon_\ell) + \Delta_\ell(E_\mathrm{cut})$, where the analytic expressions for $\Delta_0$ and $\Delta_1$ from Eqs.\ (\ref{Eq:Delta-0}) and (\ref{Eq:Delta-1}) with terms up to the order $E_\mathrm{cut}^{-3/2}$ are used. The difference $Z_1 \tilde{\Sigma}^\mathrm{c}_1(\epsilon_1) - Z_0 \tilde{\Sigma}^\mathrm{c}_0(\epsilon_0)$, which equals the contribution of the correlation part of the self-energy to the gap, is also shown together with the asymptotic formula $Z_1 \Sigma^\mathrm{c}_1(\epsilon_1) - Z_0 \Sigma^\mathrm{c}_0(\epsilon_0) + \Delta_\mathrm{gap}(E_\mathrm{cut})$. In this case, only the term proportional to $E_\mathrm{cut}^{-3/2}$ in Eq.\ (\ref{Eq:Delta-gap}) is considered, because the difference $Z_1 - Z_0 \approx 0.01$ is so small that the leading term is almost completely suppressed; the change resulting from its inclusion would not be discernible on the scale of the figure. The asymptotic convergence of the energy gap is evidently well described in this way.

In actual \textit{ab initio} calculations for real solids, the high computational cost often precludes a convergence to the desired accuracy. The contribution of the omitted high-lying unoccupied states may be approximately included within the extrapolar method\cite{Bruneval2008} or the effective-energy technique,\cite{Berger2010} but straightforward extrapolation to arrive at the limiting values would seem the most natural course of action where implementations of these schemes are not available. In practice, however, a direct extrapolation of the quasiparticle energies has only been attempted very rarely due to uncertainties about the proper asymptotic formula. In Ref.\ \onlinecite{Friedrich2011} an expression of the form
\begin{equation}
f(N) = \frac{a}{N - N_0} + b \;,
\end{equation}
where $N$ is the number of bands and $a$, $b$, and $N_0$ are fitting parameters, was employed \textit{ad hoc} to extrapolate the limiting value $b$ for the band gap of zinc oxide, whereas fitting functions with different powers of $N$ have been used in other studies.\cite{Hamada1990} As the number of bands increases proportional to $E_\mathrm{cut}^{3/2}$ at high cutoff energies, this work suggests that the form $f(N) - b \sim N^{-1} \sim E_\mathrm{cut}^{-3/2}$ guessed in Ref.\ \onlinecite{Friedrich2011} is indeed correct. The suppression of the leading order in Eq.\ (\ref{Eq:Delta-gap}) also provides an explanation why band gaps are often observed to converge faster than the individual quasiparticle energies.

\section{Conclusions\label{Sec:conclusions}}

In this paper I have described a nontrivial model system of two interacting electrons on a sphere for which the self-energy in the $GW$ approximation without self-consistency can be evaluated analytically. This corresponds to the standard approach taken in virtually all actual \textit{ab initio} calculations. As the relevant characteristics, such as the long-range Coulomb potential and the infinite Hilbert space, are the same as in real materials, the system appears better suited to explore the properties of the $GW$ approximation than previously employed analytically solvable lattice models. Indeed, the results presented here demonstrate an analogous performance as for real materials: The $GW$ approximation corrects the underestimation of the fundamental energy gap in Kohn-Sham density-functional theory and yields accurate quantitative results for low to intermediate correlation strength, but fails in the strong-correlation regime. For this reason, it suggests itself as a natural testing ground to study extensions beyond the standard $GW$ approximation that are designed to describe the self-energy of strongly correlated systems. Even if a purely analytic treatment is then no longer possible, the computational cost will be much smaller than for real materials, allowing a highly accurate evaluation without the apparent artificialities and the parameter dependence of typical lattice models, which have repeatedly been chosen to study the influence of vertex corrections and self-consistency in the past.\cite{Schindlmayr1998b,Takada2001,Romaniello2009,Sun2004,Schindlmayr1998a,Romaniello2012}

Here this system was used to study the convergence of the self-energy with respect to the number of empty states included in the spectral summations. The results not only confirm previous empirical observations that transition energies converge faster than individual quasiparticle states due to a partial error cancellation, but the asymptotic expansion also demonstrates that the gap between the highest occupied and the lowest unoccupied state approaches its limiting value with an error that is, for practical purposes, proportional to the cutoff energy to the power $-3/2$. Although a more general study of the asymptotic behavior is highly desirable, there is no indication that the dependence on the cutoff energy obtained here is due to specific details of this model. Indeed, the truncation error of the quasiparticle band gap in solids appears to exhibit the same exponent.\cite{Friedrich2011} If confirmed, this would enable practical direct extrapolation schemes with an appropriate fitting function, implying enormous potential benefits for computationally expensive \textit{ab initio} calculations. Furthermore, if some way was known to determine the relevant prefactor based on general characteristics of the material in question, then even an \textit{a posteriori} correction without the ambiguities of numerical fitting procedures would be possible. Such an \textit{a posteriori} scheme exists, for example, to extrapolate the self-energy correction of the band gap from the repeated-slab approximation with finite vacuum buffers and three-dimensional periodicity to the limit of an isolated slab, greatly accelerating the convergence of $GW$ calculations for thin films with respect to the supercell size.\cite{Freysoldt2008} In this sense, it is hoped that the present paper sparks further fruitful work along the same lines.

\begin{acknowledgments}
The author gratefully acknowledges the hospitality of the Institute for Solid State Physics at the University of Tokyo, where he performed part of this work during his tenure as a visiting professor.
\end{acknowledgments}


\begin{thebibliography}{58}
\bibitem{Hohenberg1964} P. Hohenberg and W. Kohn, Phys. Rev. \textbf{136}, B864 (1964).
\bibitem{Kohn1965} W. Kohn and L. J. Sham, Phys. Rev. \textbf{140}, A1133 (1965).
\bibitem{Foulkes2001} W. M. C. Foulkes, L. Mitas, R. J. Needs, and G. Rajagopal, Rev. Mod. Phys. \textbf{73}, 33 (2001).
\bibitem{Mahan1990} G. D. Mahan, \textit{Many-Particle Physics} (Plenum, New York, 1990).
\bibitem{Hedin1965} L. Hedin, Phys. Rev. B \textbf{139}, A796 (1965).
\bibitem{Aulbur2000} W. G. Aulbur, L. J\"onsson, and J. W. Wilkins, in \textit{Solid State Physics}, edited by H. Ehrenreich and F. Spaepen (Academic, New York, 2000), Vol. 54, p. 1.
\bibitem{Shirley1996} E. L. Shirley, Phys. Rev. B \textbf{54}, 7758 (1996).
\bibitem{Holm1998} B. Holm and U. von Barth, Phys. Rev. B \textbf{57}, 2108 (1998).
\bibitem{Schone1998} W.-D. Sch\"one and A. G. Eguiluz, Phys. Rev. Lett. \textbf{81}, 1662 (1998).
\bibitem{Schindlmayr1998b} A. Schindlmayr, T. J. Pollehn, and R. W. Godby, Phys. Rev. B \textbf{58}, 12684 (1998).
\bibitem{Takada2001} Y. Takada, Phys. Rev. Lett. \textbf{87}, 226402 (2001); H. Maebashi and Y. Takada, Phys. Rev. B \textbf{84}, 245134 (2011).
\bibitem{Ku2002} W. Ku and A. G. Eguiluz, Phys. Rev. Lett. \textbf{89}, 126401 (2002).
\bibitem{Faleev2004} S. V. Faleev, M. van Schilfgaarde, and T. Kotani, Phys. Rev. Lett. \textbf{93}, 126406 (2004).
\bibitem{Bruneval2006} F. Bruneval, N. Vast, and L. Reining, Phys. Rev. B \textbf{74}, 045102 (2006).
\bibitem{Shishkin2007a} M. Shishkin and G. Kresse, Phys. Rev. B \textbf{75}, 235102 (2007).
\bibitem{Shishkin2007b} M. Shishkin, M. Marsman, and G. Kresse, Phys. Rev. Lett. \textbf{99}, 246403 (2007).
\bibitem{Stan2009} A. Stan, N. E. Dahlen, and R. van Leeuwen, J. Chem. Phys. \textbf{130}, 114105 (2009).
\bibitem{Luttinger1960} J. M. Luttinger and J. C. Ward, Phys. Rev. \textbf{118}, 1417 (1960).
\bibitem{Baym1962} G. Baym, Phys. Rev. \textbf{112}, 1391 (1962).
\bibitem{Kotani2002} T. Kotani and M. van Schilfgaarde, Solid State Commun. \textbf{121}, 461 (2002).
\bibitem{Shishkin2006} M. Shishkin and G. Kresse, Phys. Rev. B \textbf{74}, 035101 (2006).
\bibitem{Friedrich2006} C. Friedrich, A. Schindlmayr, S. Bl\"ugel, and T. Kotani, Phys. Rev. B \textbf{74}, 045104 (2006).
\bibitem{Gomez-Abal2008} R. G\'omez-Abal, X. Li, M. Scheffler, and C. Ambrosch-Draxl, Phys. Rev. Lett. \textbf{101}, 106404 (2008).
\bibitem{Friedrich2010} C. Friedrich, S. Bl\"ugel, and A. Schindlmayr, Phys. Rev. B \textbf{81}, 125102 (2010); C. Friedrich, M. Betzinger, M. Schlipf, S. Bl\"ugel, and A. Schindlmayr, J. Phys.: Condens. Matter \textbf{24}, 293201 (2012).
\bibitem{Li2012} X.-Z. Li, R. G\'omez-Abal, H. Jiang, C. Ambrosch-Draxl, and M. Scheffler, New J. Phys. \textbf{14} 023006 (2012).
\bibitem{Freysoldt2007} C. Freysoldt, P. Eggert, P. Rinke, A. Schindlmayr, R. W. Godby, and M. Scheffler, Comput. Phys. Commun. \textbf{176}, 1 (2007).
\bibitem{Friedrich2009} C. Friedrich, A. Schindlmayr, and S. Bl\"ugel, Comput. Phys. Commun. \textbf{180}, 347 (2009).
\bibitem{vanSchilfgaarde2006} M. van Schilfgaarde, T. Kotani, and S. V. Faleev, Phys. Rev. B \textbf{74}, 245125 (2006).
\bibitem{Tiago2004} M. L. Tiago, S. Ismail-Beigi, and S. G. Louie, Phys. Rev. B \textbf{69}, 125212 (2004).
\bibitem{Delaney2004} K. Delaney, P. Garc\'{\i}a-Gonz\'alez, A. Rubio, P. Rinke, and R. W. Godby, Phys. Rev. Lett. \textbf{93}, 249701 (2004); W. Ku and A. G. Eguiluz, \textit{ibid.} \textbf{93}, 249702 (2004).
\bibitem{Bruneval2008} F. Bruneval and X. Gonze, Phys. Rev. B \textbf{78}, 085125 (2008).
\bibitem{Shih2010} B.-C. Shih, Y. Xue, P. Zhang, M. L. Cohen, and S. G. Louie, Phys. Rev. Lett. \textbf{105}, 146401 (2010).
\bibitem{Stankovski2011} M. Stankovski, G. Antonius, D. Waroquiers, A. Miglio, H. Dixit, K. Sankaran, M. Giantomassi, X. Gonze, M. C\^ot\'e, and G.-M. Rignanese, Phys. Rev. B \textbf{84}, 241201(R) (2011).
\bibitem{Friedrich2011} C. Friedrich, M. C. M\"uller, and S. Bl\"ugel, Phys. Rev. B \textbf{83}, 081101(R) (2011); \textbf{84}, 039906(E) (2011).
\bibitem{Steinbeck2000} L. Steinbeck, A. Rubio, L. Reining, M. Torrent, I. D. White, and R.W. Godby, Comput. Phys. Commun. \textbf{125}, 105 (2000).
\bibitem{Umari2010} P. Umari, G. Stenuit, and S. Baroni, Phys. Rev. B \textbf{81}, 115104 (2010).
\bibitem{Berger2010} J. A. Berger, L. Reining, and F. Sottile, Phys. Rev. B \textbf{82}, 041103(R) (2010); \textbf{85}, 085126 (2012).
\bibitem{Giustino2010} F. Giustino, M. L. Cohen, and S. G. Louie, Phys. Rev. B \textbf{81}, 115105 (2010).
\bibitem{Mahan1989} G. D. Mahan and B. E. Sernelius, Phys. Rev. Lett. \textbf{62}, 2718 (1989).
\bibitem{Holm1997} B. Holm and F. Aryasetiawan, Phys. Rev. B \textbf{56}, 12825 (1997).
\bibitem{Schindlmayr1997} A. Schindlmayr, Phys. Rev. B \textbf{56}, 3528 (1997).
\bibitem{Schindlmayr2001} A. Schindlmayr, P. Garc\'{\i}a-Gonz\'alez, and R. W. Godby, Phys. Rev. B \textbf{64}, 235106 (2001).
\bibitem{Romaniello2009}  P. Romaniello, S. Guyot, and L. Reining, J. Chem. Phys. \textbf{131}, 154111 (2009).
\bibitem{Verdozzi1995} C. Verdozzi, R. W. Godby, and S. Holloway, Phys. Rev. Lett. \textbf{74}, 2327 (1995).
\bibitem{Pollehn1998} T. J. Pollehn, A. Schindlmayr, and R. W. Godby, J. Phys.: Condens. Matter \textbf{10}, 1273 (1998).
\bibitem{Sun2004} P. Sun and G. Kotliar, Phys. Rev. Lett. \textbf{92}, 196402 (2004).
\bibitem{Kaasbjerg2010} K. Kaasbjerg and K. S. Thygesen, Phys. Rev. B \textbf{81}, 085102 (2010).
\bibitem{Hedin1999} L. Hedin, J. Phys.: Condens. Matter \textbf{11}, R489 (1999).
\bibitem{Loos2009a} P.-F. Loos and P. M. W. Gill, Phys. Rev. A \textbf{79}, 062517 (2009).
\bibitem{Loos2009b} P.-F. Loos and P. M. W. Gill, Phys. Rev. Lett. \textbf{103}, 123008 (2009).
\bibitem{Levy1984} M. Levy, J. P. Perdew, and V. Sahni, Phys. Rev. A \textbf{30}, 2745 (1984).
\bibitem{Gaunt1929} J. A. Gaunt, Philos. Trans. R. Soc., A \textbf{228}, 151 (1929).
\bibitem{Messiah1962} A. Messiah, \textit{Quantum Mechanics} (North-Holland, Amsterdam, 1962), Vol. 2.
\bibitem{Erdelyi1953} \textit{Higher Transcendental Functions}, edited by A. Erd\'elyi (McGraw-Hill, New York, 1953), Vol. 1.
\bibitem{Hamada1990} N. Hamada, M. Hwang, and A. J. Freeman, Phys. Rev. B \textbf{41}, 3620 (1990).
\bibitem{Schindlmayr1998a} A. Schindlmayr and R. W. Godby, Phys. Rev. Lett. \textbf{80}, 1702 (1998).
\bibitem{Romaniello2012} P. Romaniello, F. Bechstedt, and L. Reining, Phys. Rev. B \textbf{85}, 155131 (2012).
\bibitem{Freysoldt2008} C. Freysoldt, P. Eggert, P. Rinke, A. Schindlmayr, and M. Scheffler, Phys. Rev. B \textbf{77}, 235428 (2008).
\end{thebibliography}
\end{document}